\documentclass[12pt]{article}
\usepackage{amsmath}
\usepackage[unicode,bookmarks,bookmarksopen,bookmarksopenlevel=2,colorlinks,linkcolor=blue,citecolor=green]{hyperref}

\newtheorem{theorem}{Theorem}

\begin{document}

\title{\textbf{On Local Description \\ of Two-Dimensional Geodesic Flows \\ with a
Polynomial First Integral}}
\author{Maxim V. Pavlov$^{1,2}$, Sergey P. Tsarev$^{3}$ \\
$^{1}$Sector of Mathematical Physics,\\
Lebedev Physical Institute of Russian Academy of Sciences,\\
Leninskij Prospekt 53, 119991 Moscow, Russia\\
$^{2}$Department of Applied Mathematics,\\
National Research Nuclear University MEPHI,\\
Kashirskoe Shosse 31, 115409 Moscow, Russia\\
$^{3}$Siberian Federal University, \\
Institute of Space and Information Technologies,\\
26 Kirenski str., ULK-311, \\
Krasnoyarsk, 660074 Russia \\
}
\date{\today}
\maketitle

\begin{flushright}
\textit{dedicated to the 55th birthday \\
of our friend E.V. Ferapontov}
\end{flushright}

\begin{abstract}
In this paper we construct multiparametric families of two dimensional
metrics with polynomial first integral. Such integrable geodesic flows are
described by solutions of some semi-Hamiltonian hydrodynamic type system. We
find infinitely many conservation laws and commuting flows for this system.
This procedure allows us to present infinitely many particular metrics by
the generalized hodograph method.
\end{abstract}
\bigskip

\textit{keywords}: Geodesic flows, integrability,
hydrodynamic type system, Generalized Hodograph Method.

\bigskip



\newpage

\tableofcontents

\section{Introduction}

\label{sec-intro}

The problem of integration of geodesic flows on a two-dimensional surface
appeared in classical mechanics very early and was extensively investigated
in the XIX-th century focusing mostly on local approach. The XX-th century
was more centered on global behavior, problems of local description of
trajectories were not that much popular. An overview of both aspects may be
found in numerous sources, we give here only two references \cite{BolMatFom,
BolFom}. Partially this loss of interest to the local problem was not only
due to the importance of the global problems; it seems reasonable to ascribe
this loss of interest to absence of new ideas for the local integrability
problem. The situation, in our opinion, has changed in the last decades
after publications \cite{deryabin1997}, \cite{Bialy} where the authors had
remarked that the equations for the coefficients of first integrals
polynomial in momenta for two specific low-dimensional cases belong to the
class of diagonalizable hydrodynamic type systems integrable by
differential-geometric means; the appropriate theory for such nonlinear
systems of PDEs was developed in the very end of the XX-th century (cf. \cite%
{dn, Tsar}). In \cite{classmech}, developing the preliminary results of \cite%
{deryabin1997}, we demonstrated how to apply the techniques of integrable
hydrodynamic type systems to the case of so called
one-and-a-half-dimensional systems (one-dimensional mechanical systems with
the potential depending on time). Below we investigate (using a bit more
sophisticated technologies) the problem of local description of
two-dimensional Riemannian metrics with geodesic flows possessing a
polynomial first integral of arbitrary high degree.

In \cite{Bialy} the $N$ component hydrodynamic type system
\begin{equation}
a_{t}^{0}=a^{1}a_{x}^{N-1},\text{ \ \ }%
a_{t}^{k}=a^{N-1}a_{x}^{k-1}+[(k+1)a^{k+1}-(N+1-k)a^{k-1}]a_{x}^{N-1},
\label{eq:a_i}
\end{equation}%
where $k=1,\ldots ,N-1$ and $a^{N}\equiv 1$, was derived 
as the system for the coefficients $a^{k}(x,t)$ of a (global smooth)
polynomial first integral\footnote{The factor $(-1)^k$ was omitted in \cite{Bialy}.}
\begin{equation}
F(x,t,p_{1},p_{2})=\overset{N}{\underset{k=0}{\sum }}\frac{(-1)^k a^{k}}{g^{N-k}}%
p_{1}^{N-k}p_{2}^{k},  \label{firstintegral}
\end{equation}%
for metric in semi-geodesic coordinates%
\begin{equation}
ds^{2}=g^{2}(x,t)dt^{2}+dx^{2}  \label{metric_g}
\end{equation}%
on a 2-dimensional torus, where $g\equiv a^{N-1}$. As shown in \cite{Bialy},
any smooth metric on a 2-dimensional torus possessing a polynomial first
integral can be reduced to such a form. The metric (\ref{metric_g}) corresponds to
the Hamiltonian $H(x,t,p_1,p_2) = \frac{1}{2}(p_1^{2}/g^{2}(x,t)+p_2^{2}$); the system (\ref{eq:a_i})
is equivalent to
\begin{equation*}
    \{F,H\} = \frac{\partial F}{\partial p_1} \frac{\partial H}{\partial t}
              - \frac{\partial F}{\partial t} \frac{\partial H}{\partial p_1}
              + \frac{\partial F}{\partial p_2} \frac{\partial H}{\partial x}
              - \frac{\partial F}{\partial x} \frac{\partial H}{\partial p_2} = 0.
\end{equation*}
The theorem proved in \cite{Bialy} states that the hydrodynamic type system (%
\ref{eq:a_i}) is diagonalizable (i.e. possesses a complete set of Riemann
invariants) and semi-Hamiltonian so may be integrated by the Generalized
Hodograph Method (cf. \cite{Tsar}). However the authors of \cite{Bialy} did
not give a constructive description of the necessary complete set of
hydrodynamic conservation laws and commuting flows for (\ref{eq:a_i}).

In this paper 
we construct $N$ infinite series of conservation laws and commuting flows.
Thus one can construct a rich multiparametric family of particular solutions
to (\ref{eq:a_i}) by the Generalized Hodograph Method as described below in
Sections~\ref{sec:integrability},~\ref{sec:GenHod}. Our interest is focused
on local properties of the system (\ref{eq:a_i}) and the respective
coefficient $g(x,t)$ in (\ref{metric_g}).

The structure of the paper is as follows. In Section~\ref{sec:integrability}
we discuss the semi-Hamiltonian property of hydrodynamic type system (\ref%
{eq:a_i}). We construct a generating function of conservation laws and find
the equation of the associated Riemann surface. We rewrite hydrodynamic type
system (\ref{eq:a_i}) in a diagonal form. In Section~\ref{sec:char_vel} we
rewrite hydrodynamic type system (\ref{eq:a_i}) via characteristic
velocities and derive analogues of the L\"{o}wner equations and the
Gibbons--Tsarev system.
Also we remark that in the two component case hydrodynamic type system (\ref%
{eq:a_i}) is nothing but the simplest two-component linearly degenerate
hydrodynamic type system (which can be written in appropriate field
variables in the form $u_{t}=vu_{x},v_{t}=uv_{x}$), whose general solution
can be presented in implicit form with explicit dependence on two arbitrary
functions of a single variable. Such a system has global solutions. This
result conforms to the well-known fact that geodesic flows with quadratic
first integrals can be integrated by separation of variables. In Section~\ref%
{sec:canon} we introduce new field variables $b^{k}$
and show how the equation of a Riemann surface rewritten in these field
variables $b^{k}$ helps in constructing $N$ infinite series of conservation
law densities. In Section~\ref{sec:chain} we introduce two hydrodynamic
chains as infinite sets of equations compatible with hydrodynamic type
system (\ref{eq:a_i}). 
This procedure allows us to construct infinitely many conservation laws (the
so called Kruskal series) in a compact form. In Section~\ref{sec:comm_flows}
we present
the way to construct infinitely many higher commuting flows and infinitely
many associated conservation laws. In Section~\ref{sec:GenHod} we adopted
the Generalized Hodograph Method for construction of a rich
infinite-parametric family of 
particular solutions for the case of hydrodynamic type system (\ref{eq:a_i}%
). Finally in Conclusion (Section~\ref{sec:final}) we discuss the problem of
completeness of the constructed series of conservation laws and briefly
expose further perspectives of integrability of two-dimensional geodesic
flows.

\section{Semi-Hamiltonian Systems and their Integration}

\label{sec:integrability}

Integrability of a diagonalizable hydrodynamic type system (similar to (\ref%
{eq:a_i})) means that such a system possesses $N$ Riemann invariants $%
\mathbf{r} = (r^1, \ldots , r^N)$, has infinitely many hydrodynamic
conservation laws and locally any solution can be constructed by the
Generalized Hodograph Method:
\begin{equation}  \label{eq:GHM}
w_i(r^1, \ldots , r^N) = v_i(r^1, \ldots , r^N) \cdot t + x, \qquad i = 1,
\ldots , N,
\end{equation}
where $v_i(\mathbf{r})$ are the characteristic velocities of the system (\ref%
{eq:a_i}) in the diagonal form
\begin{equation}  \label{eq:diag_v}
r^i_t = v_i(\mathbf{r}) r^i_x, \qquad i = 1, \ldots , N,
\end{equation}
and $w_i(\mathbf{r})$ are the coefficients of commuting with (\ref{eq:a_i})
flows
\begin{equation}  \label{eq:comm_w}
r^i_\tau = w_i(\mathbf{r}) r^i_x, \qquad i = 1, \ldots , N,
\end{equation}
(no summation over repeated indices is assumed anywhere in this paper).

Such systems were called \textquotedblleft
semi-Hamiltonian\textquotedblright in \cite{Tsar} where the detailed
exposition of the respective differential-geometric theory was given.

In this Section we present the equation of a Riemann surface $\lambda (q,%
\mathbf{a})$ associated with the hydrodynamic type system (\ref{eq:a_i}).
Its branch points $r^{i}=\lambda |_{q=q_{i}}$, where $q_{i}$ are solutions
of the algebraic equation $\lambda _{q}=0$, are the Riemann invariants of (%
\ref{eq:a_i}). Also we briefly remind how to find a rich infinite family of
conservation laws and commuting flows (\ref{eq:comm_w}) for (\ref{eq:diag_v}%
).

\textbf{Theorem} \cite{Bialy}: \textit{Hydrodynamic type system }(\ref%
{eq:a_i}) \textit{is semi-Hamiltonian}.

In this Section we give an alternative proof of this Theorem obtaining the
associated Riemann surface playing the key role in our construction of the
explicit formulas for the conservation laws and commuting flows for (\ref%
{eq:a_i}).

\textbf{Proof}: According to \cite{Bialy} hydrodynamic type system (\ref%
{eq:a_i}) can be equally derived for the coefficients $a^{k}(x,t)$ of a
polynomial first integral\footnote{The corresponding formula in \cite{Bialy} contains misprints.}
\begin{equation}\label{eq:F_tilda}
\tilde{F}(x,t,p)=\sum_{k=0}^{N} (-1)^k a^{k}(x,t)p^{k}(1-p^{2})^{\frac{N-k}{2}}
\end{equation}%
of Hamilton's equations\footnote{%
This means that geodesic flows can be written either as stationary
Hamilton's equations with two degree of freedom or alternatively as \textit{%
non-stationary} Hamilton's equations with one-and-a-half degree of freedom.
Here the \textquotedblleft prime\textquotedblright\ means the derivative
with respect to \textquotedblleft $t$\textquotedblright .}%
\begin{equation}
x^{\prime }=\frac{\partial \tilde{H}}{\partial p},\text{ \ }p^{\prime }=-\frac{%
\partial \tilde{H}}{\partial x},  \label{hamilton}
\end{equation}%
where the \textit{effective} Hamiltonian function is%
\begin{equation*}
\tilde{H}(x,t,p)=-a^{N-1}(x,t)\sqrt{1-p^{2}}. 
\end{equation*}%
This means that in fact we have a one-dimensional mechanical system with
Hamiltonian depending explicitly on the \textquotedblleft
time\textquotedblright\ $t$. Such systems were called \textquotedblleft
1.5-dimensional\textquotedblright\ in \cite{kozlov}. An integrable subclass
of 1.5-dimensional systems was studied in \cite{classmech} where the general
technology of hydrodynamic reductions of integrable nonlinear hydrodynamic
chains was used for explicit description of various integrable
1.5-dimensional cases. In the present paper we develop a more general
approach for the case studied.

In this 1.5-dimensional setting the corresponding Liouville equation%
\footnote{%
The Poisson bracket is standard $\{f,H\}=f_{p}H_{x}-f_{x}H_{p}$. We remind
that $f(x,t,p)$ is a first integral for the given Hamilton's equations,
for example $f= \tilde F$.}%
\begin{equation*}
f_{t}=\{f,\tilde{H}\} =
\frac{\partial f}{\partial p} \frac{\partial \tilde{H}}{\partial x}
              - \frac{\partial f}{\partial x} \frac{\partial \tilde{H}}{\partial p}
\end{equation*}%
takes the form%
\begin{equation}
f_{t}=a^{N-1}qf_{x}+(1+q^{2})f_{q}a_{x}^{N-1},  \label{eq:f_t}
\end{equation}%
where instead of the variable $p$ we introduce the auxiliary variable $q$
connected with $p$ via the point transformation%
\begin{equation}
q=-\frac{p}{\sqrt{1-p^{2}}},\text{ \ \ }p=-\frac{q}{\sqrt{1+q^{2}}}.
\label{eq:q2p}
\end{equation}%
Substitution of a general ansatz $f(x,t,p)=\lambda (q,\mathbf{a}(x,t))$ with
a set of some (formally unspecified) field variables $\mathbf{a}\equiv
(a^{0}(x,t),\ldots ,a^{N-1}(x,t))$ into (\ref{eq:f_t}) leads to an
overdetermined compatible systems of equations on $\partial \lambda
/\partial a^{i}$, $\partial \lambda /\partial q$ if $a^{k}(x,t)$ are
solutions of hydrodynamic type system (\ref{eq:a_i}). Now $\lambda (q,%
\mathbf{a}(x,t))$ can be found explicitly:
\begin{equation}
\lambda (q,\mathbf{a})=(1+q^{2})^{-N/2}\left( q^{N}+\overset{N-1}{\underset{%
k=0}{\sum }}q^{k}a^{k}\right) .  \label{eq:Riemann}
\end{equation}%
In fact (\ref{eq:Riemann}) can be obtained by a direct substitution of (\ref{eq:q2p})
into  (\ref{eq:F_tilda}).
Thus a Riemann surface with parameters $(\lambda ,q)$ used below for the
explicit constructions of the conservations laws and commuting flows is
defined. And vice versa, substitution of (\ref{eq:Riemann}) into (\ref%
{eq:f_t}) and splitting w.r.t. $q$ leads to (\ref{eq:a_i}). In a generic
case the algebraic equation $\lambda _{q}=0$ for $\lambda $ of the form (\ref%
{eq:Riemann}), i.e.
\begin{equation}
\overset{N-1}{\underset{k=0}{\sum }}[(N-k)q^{k+1}-kq^{k-1}]a^{k}=Nq^{N-1},
\label{eq:roots}
\end{equation}%
has $N$ simple distinct roots $q_{i}=q_{i}(\mathbf{a})$. Since $\lambda
_{q}=0$ in these points $q_{i}=q_{i}(\mathbf{a}(x,t))$, Liouville equation (%
\ref{eq:f_t}) leads to the hydrodynamic type system (\ref{eq:a_i}), written
in the diagonal form%
\begin{equation}
r_{t}^{i}=a^{N-1}q_{i}r_{x}^{i},  \label{eq:inv_rim}
\end{equation}%
where the Riemann invariants $r^{i}=\lambda |_{q=q_{i}}$ are determined by (%
\ref{eq:Riemann}), i.e.
\begin{equation}
r^{i}(\mathbf{a})=\lambda (q_{i},\mathbf{a})=(1+(q_{i})^{2})^{-N/2}\left(
(q_{i})^{N}+\overset{N-1}{\underset{m=0}{\sum }}(q_{i})^{m}a^{m}\right) ,
\label{eq:inv_rimm}
\end{equation}%
It is easy to see that the constructed $r^{i}(\mathbf{a})$ are functionally
independent since the characteristic velocities in (\ref{eq:inv_rim}) are
distinct. (Functional independence of the velocities $v_{i}=a^{N-1}q_{i}$
will be shown below in Section~\ref{sec:char_vel}.)

Under the functional inversion transformation $\lambda = \lambda(q,\mathbf{a}%
)\rightarrow q = q(\lambda , \mathbf{a})$ the linear equation (\ref{eq:f_t})
transforms to 
\begin{equation}
q_{t}=a^{N-1}qq_{x}-(1+q^{2})a_{x}^{N-1},  \label{eq:qus}
\end{equation}%
which is equivalent to the equation\footnote{%
Under the potential substitution $p=S_{x}$ this equation reduces to the
Hamilton--Jacobi equation $S_{t}=a^{N-1}\sqrt{1-(S_{x})^{2}}$ for the
mechanical system with the specified Hamiltonian $H(x,t,p)$.}%
\begin{equation}
p_{t}=\left( a^{N-1}\sqrt{1-p^{2}}\right) _{x}  \label{eq:p_t}
\end{equation}%
for the so-called generating function of conservation law densities $p =
p(\lambda ,a^{0},a^{1}, \ldots ,a^{N-1})$
where $\lambda $ is a parameter. Infinitely many conservation laws can be
found directly from the equation of Riemann surface (\ref{eq:Riemann})
expanding the inverse function $q(\lambda ,\mathbf{a} )$ (as $q \rightarrow
\infty $, while $\lambda \rightarrow 1$) w.r.t. the powers of $(\lambda - 1)$
and substituting into the second relationship of (\ref{eq:q2p}) and then
into (\ref{eq:p_t}).\footnote{%
This remarkable class of conservation laws is usually called Kruskal series.
We describe its construction in more detail below in Section~\ref{sec:chain}.%
} Since coefficients of the resulting expansion of $p(\lambda ,\mathbf{a})$
are conservation law densities (see (\ref{eq:p_t})), and the hydrodynamic
type system (\ref{eq:a_i}) is diagonalizable (see (\ref{eq:inv_rim})), we
conclude that (\ref{eq:a_i}) is semi-Hamiltonian. The Theorem is proved.

The next important step in integration of semi-Hamiltonian system (\ref%
{eq:a_i}) is construction of the necessary complete set of $w_i(\mathbf{r})$
(depending on $N$ functions of one variable, cf. \cite{Tsar}) in order to be
able to construct any solution of (\ref{eq:a_i}) locally using the
Generalized Hodograph formula (\ref{eq:GHM}). This problem will be discussed
in Sections~\ref{sec:comm_flows},~\ref{sec:GenHod}. In the following three
Sections we consider different forms of the system (\ref{eq:a_i}) and the
associated Riemann surface (\ref{eq:Riemann}) in more detail in order to
expose the necessary techniques.

\section{More on Characteristic Velocities and Riemann Invariants.
Triviality of the Quadratic Case ($N=2$).}

\label{sec:char_vel}

Since the field variables $a^{k}$ in (\ref{eq:a_i}) are connected with the
characteristic velocities (the eigenvalues of the system matrix)
algebraically (see (\ref{eq:roots}) and (\ref{eq:inv_rim})), one could try
to find explicitly the dependence of the functions $a^{k}$ on the
characteristic velocities $v_i(\mathbf{a}) = a^{N-1}q_i(\mathbf{a})$, if $%
v_i(\mathbf{a})$ are functionally independent. Thus automatically the
hydrodynamic type system studied in this paper would be written in a
symmetric form.

Indeed, (\ref{eq:qus}) yields directly the same hydrodynamic type system (%
\ref{eq:a_i}) written as
\begin{equation}
(q_{k})_{t}=a^{N-1}q_{k}(q_{k})_{x}-(1+q_{k}^{2})a_{x}^{N-1},  \label{quq}
\end{equation}%
where we just replaced $q$ by the corresponding roots $q_{k}$.


\textbf{Remark}: This replacement ($q\rightarrow q_{k}$) is not so obvious,
because roots $q_{k}$ of algebraic equation determined by the condition $%
\lambda _{q}=0$ do not depend on the parameter $\lambda $, while the
function $q$ depends on it. In fact $q_k(\mathbf{a}) = q(\lambda,\mathbf{a}%
)|_{\lambda = r^k(\mathbf{a})}$. For this reason we give here a precise
derivation of the above equations. 
The general theory of hydrodynamic reductions developed in \cite{GT} implies
that $q(\lambda ,\mathbf{a(r)})$ satisfies the overdetermined system (the so
called L\"{o}wner type equation)%
\begin{equation}
\partial _{i}q=\frac{1+q^{2}}{q_{i}-q}\partial _{i}A, \qquad \partial _{i} =
{\partial}/{\partial r^{i}},  \label{lowner}
\end{equation}%
where we denoted $A=-\ln a^{N-1}$. These equations are in fact the equation (%
\ref{eq:qus}) for $q(\lambda ,\mathbf{a(r)})$ after substitution of (\ref%
{eq:inv_rim}) and splitting w.r.t. $r^{i}_x$. Compatibility conditions $%
\partial _{k}(\partial _{i}q)=\partial _{i}(\partial _{k}q)$ for (\ref%
{lowner}) yield the so called Gibbons--Tsarev type system%
\begin{equation}
\partial _{i}q_{k}=\frac{1+q_{k}^{2}}{q_{i}-q_{k}}\partial _{i}A,\qquad
\partial _{i} \partial _{k}A=2\frac{1+q_{i}q_{k}}{(q_{i}-q_{k})^{2}}\partial
_{i}A \partial _{k}A, \qquad i\neq k.  \label{gt}
\end{equation}%
One can immediately verify that if $q_{k}$ are determined by (\ref{gt}),
then (\ref{quq}) is fulfilled. This computation shows that the first (left)
part of the Gibbons--Tsarev type system can be formally obtained from (\ref%
{lowner}) by replacement $q\rightarrow q_{k}$ as well as (\ref{quq}) from (%
\ref{eq:qus}).

Taking into account the relationship (see (\ref{eq:inv_rim})) between
characteristic velocities $v_{k}=a^{N-1}q_{k}$ and roots $q_{k}$, the
hydrodynamic type system (\ref{quq}) takes the form%
\begin{equation}
(v_{k})_{t}=v_{k}v_{x}^{k}+v_{k}a_{x}^{N-2}-\frac{%
2(v_{k})^{2}+(2a^{N-2}-N)v_{k}+(a^{N-1})^{2}}{a^{N-1}}a_{x}^{N-1},
\label{velo}
\end{equation}%
where $a^{N-2}$ and $a^{N-1}$ can be found from (\ref{eq:roots}), written in
the form
\begin{equation*}
\overset{N-1}{\underset{k=0}{\sum }}\left( (N-k)\frac{(v_{j})^{k+1}}{%
(a^{N-1})^{k+1}}-k\frac{(v_{j})^{k-1}}{(a^{N-1})^{k-1}}\right) a^{k}=N\frac{%
(v_{j})^{N-1}}{(a^{N-1})^{N-1}}.
\end{equation*}%
$j=1,\ldots ,N$. This is a linear algebraic system on unknown functions $%
a^{k}(\mathbf{v})$, $k=0,\ldots ,(N-2)$, except the latest coefficient $%
a^{N-1}(\mathbf{v})$. One can find, for instance, that%
\begin{equation*}
a^{N-2}=\frac{N}{2}-\frac{1}{2}\overset{N}{\underset{m=1}{\sum }}v_{m},
\end{equation*}%
while $a^{N-1}(\mathbf{v})$ satisfies different algebraic equations for
different $N$. For instance, if $N=2$, then $(a^{1})^{2}=-V_{1}$; if $N=3$,
then%
\begin{equation*}
(a^{2})^{2}=\frac{2V_{3}}{3-V_{1}};
\end{equation*}%
if $N=4$, then%
\begin{equation*}
(a^{3})^{4}+\frac{1}{3}V_{2}(a^{3})^{2}+V_{4}=0;
\end{equation*}%
if $N=5$, then%
\begin{equation*}
(a^{4})^{4}-2\frac{V_{3}}{3(5-V_{1})}(a^{4})^{2}-\frac{8V_{5}}{3(5-V_{1})}=0,
\end{equation*}%
etc. Here $V_{k}$ are the elementary symmetric polynomials of $v_{j}$ of
degree $k$ found by the Vieta Theorem from the formal equation
\begin{equation*}
v^{k}-V_{1}v^{k-1}+V_{2}v^{k-2}-V_{3}v^{k-3}+\ldots =0,
\end{equation*}%
so $V_{1}=\sum v_{m},V_{2}=\frac{1}{2}[(\sum v_{m})^{2}-\sum v_{m}^{2}],$
\ldots

Taking into account (\ref{eq:inv_rimm}) we see that all Riemann invariants $%
r^{k}$ can be explicitly expressed via the characteristic velocities $v_{m}$%
. If $N=2$, we get
\begin{equation}
r^{1}=1-\frac{1}{2}v_{2},\text{ \ \ }r^{2}=1-\frac{1}{2}v_{1},  \label{rim2}
\end{equation}%
where the corresponding equation of the Riemann surface becomes%
\begin{equation*}
\lambda (q,\mathbf{v})=1-V_{2}\frac{v-\frac{1}{2}V_{1}}{v^{2}-V_{2}};
\end{equation*}%
if $N=3$, then%
\begin{equation*}
r^{k}=\frac{(v_{k})^{3}+\frac{2V_{3}}{3-V_{1}}(v_{k})^{2}+V_{3}v_{k}+\frac{1%
}{3}\frac{2V_{3}}{3-V_{1}}\left( V_{2}+\frac{4V_{3}}{3-V_{1}}\right) }{%
\left( (v_{k})^{2}+\frac{2V_{3}}{3-V_{1}}\right) ^{3/2}};
\end{equation*}%
%
%
%
%
%
%
if $N=4$, then%
\begin{equation*}
r^{k}=1+\frac{(a^{3})^{2}(v_{k})^{3}-\frac{1}{2}%
V_{1}(a^{3})^{2}(v_{k})^{2}-V_{4}v_{k}+\frac{1}{12}V_{1}V_{2}(a^{3})^{2}-%
\frac{1}{4}(a^{3})^{2}V_{3}+\frac{1}{4}V_{1}V_{4}}{%
((v_{k})^{2}+(a^{3})^{2})^{2}}.
\end{equation*}%
%
%
%
%
%
%

However, inverse formulas for $v_{k}(\mathbf{r})$ are much more complicated.
Only if $N=2$, the inversion is simple (see (\ref{rim2})):
\begin{equation*}
v_{2}=2-2r^{1},\text{ \ \ }v_{1}=2-2r^{2}.
\end{equation*}%
Thus, hydrodynamic type system (\ref{eq:a_i}) in the two component case%
\begin{equation}
a_{t}^{0}=a^{1}a_{x}^{1},\text{ \ \ }%
a_{t}^{1}=a^{1}a_{x}^{0}+2(1-a^{0})a_{x}^{1}  \label{two}
\end{equation}%
is linearly degenerate:%
\begin{equation}
r_{t}^{1}=2(1-r^{2})r_{x}^{1},\text{ \ }r_{t}^{2}=2(1-r^{1})r_{x}^{2},
\label{dva}
\end{equation}%
where%
\begin{equation*}
a^{0}=r^{1}+r^{2}-1,\text{ \ \ }(a^{1})^{2}=-4(r^{1}-1)(r^{2}-1).
\end{equation*}

\textbf{Remark}: Algebraic equation (\ref{eq:roots}) determines $N$ roots $%
q_{k}(\mathbf{a})$. However, one can see that inverse expressions $a^{k}(%
\mathbf{q})$ cannot be found if $N$ is even due to degeneracy of the mapping
$(a^0 , \ldots , a^{N-1}) \longrightarrow (q_1, \ldots , q_{N})$ for even $N$%
. But if $N$ is odd, the inverse expressions $a^{k}(\mathbf{q})$ can be
found. For instance, if $N=3$,
\begin{equation*}
a^{1}=\frac{3Q_{3}}{Q_{1}+2Q_{3}},\text{ \ }a^{0} =\frac{Q_{2}+2}{%
Q_{1}+2Q_{3}},\text{ \ }a^{2}=\frac{3}{Q_{1}+2Q_{3}},
\end{equation*}%
where $%
Q_{1}=q_{1}+q_{2}+q_{3},Q_{2}=q_{1}q_{2}+q_{1}q_{3}+q_{2}q_{3},Q_{3}=q_{1}q_{2}q_{3}
$. The corresponding expression for the equation of the Riemann surface (\ref%
{eq:Riemann}) is
\begin{equation*}
\lambda (q,\mathbf{q})=\frac{(1+q^{2})^{-3/2}}{Q_{1}+2Q_{3}}%
[(Q_{1}+2Q_{3})q^{3}+3q^{2}+3Q_{3}q+Q_{2}+2].
\end{equation*}

\subsection{Linearly Degenerate Case ($N=2$)}

Two component hydrodynamic type system (\ref{two}) corresponding to
quadratic first integrals (\ref{firstintegral}) of geodesics on a
two-dimensional surface is linearly degenerate (see (\ref{dva})), that is $%
\partial v_{1}/\partial r^{1}=0$ and $\partial v_{2}/\partial r^{2}=0$. Its
general solution $r^{1}(x,t)$, $r^{2}(x,t)$ depending on two arbitrary
functions $\beta (u)$ and $\gamma (v)$ of one variable is well known (cf.,
for example \cite{RozYa}) and may be presented in implicit form:
\begin{equation*}
t=\beta ^{\prime }(u)+\gamma ^{\prime }(v),\text{ \ }x=\beta (u)-u\beta
^{\prime }(u)+\gamma (v)-v\gamma ^{\prime }(v),
\end{equation*}%
where (for simplicity) we denoted $u=2(1-r^{1}),v=2(1-r^{2})$.

The fact of complete integrability in the case of quadratic first integrals
is also well known, its exposition can be found for example in \cite%
{BolMatFom} and \cite[Ch.~11]{BolFom}. Actually the standard exposition of
this case (for the isothermic form $ds^{2}=(f(u)+g(v))(du^{2}+dv^{2})$ of
the metric) in the references given above is equivalent to our result for
the metric in semi-geodesic coordinates (\ref{metric_g}).

\section{Symmetric Field Variables and $N$ Principal Series of Conservation
Laws}

\label{sec:canon}

In the previous Section the hydrodynamic type system (\ref{eq:a_i}) was
written in symmetric form (\ref{velo}). However such a symmetric form is non
unique. Let us introduce the roots $b^{k}$ of the polynomial
\begin{equation*}
q^{N}+\overset{N-1}{\underset{k=0}{\sum }}q^{k}a^{k}=\overset{N}{\underset{%
k=1}{\prod }}(q-b^{k}), 
\end{equation*}%
so all field variables $a^{k}$ become elementary symmetric polynomials of
new field variables $b^{n}$. For instance,%
\begin{equation}
a^{N-1}=-\overset{N}{\underset{k=1}{\sum }}b^{k}.  \label{suma}
\end{equation}

Thus, the equation of the Riemann surface (\ref{eq:Riemann}) takes the form%
\begin{equation}
\lambda =(1+q^{2})^{-N/2}\overset{N}{\underset{k=1}{\prod }}(q-b^{k}).
\label{riman}
\end{equation}%
After substitution of this expression into (\ref{eq:f_t}) the hydrodynamic
type system (\ref{eq:a_i}) reduces to another simple symmetric form
\begin{equation}
b_{t}^{k}=(1+(b^{k})^{2})\overset{N}{\underset{m=1}{\sum }}b_{x}^{m}-\left(
\overset{N}{\underset{n=1}{\sum }}b^{n}\right) b^{k}b_{x}^{k}.
\label{eq:b_t}
\end{equation}
Indeed, hydrodynamic type system (\ref{eq:b_t}) can be derived in three
steps:

1. compute the partial derivatives of $\ln \lambda $ with respect to the
independent variables $x,t,q$:
\begin{equation*}
(\ln \lambda )_{x}=-\overset{N}{\underset{m=1}{\sum }}\frac{b_{x}^{m}}{%
q-b^{m}},\text{ \ }(\ln \lambda )_{t}=-\overset{N}{\underset{m=1}{\sum }}%
\frac{b_{t}^{m}}{q-b^{m}},\text{ \ }(\ln \lambda )_{q}=-\frac{Nq}{1+q^{2}}+%
\overset{N}{\underset{m=1}{\sum }}\frac{1}{q-b^{m}};
\end{equation*}

2. substitution of these derivatives into (\ref{eq:f_t}) leads to%
\begin{equation*}
\overset{N}{\underset{m=1}{\sum }}\frac{b_{t}^{m}}{q-b^{m}}=a^{N-1}q\overset{%
N}{\underset{m=1}{\sum }}\frac{b_{x}^{m}}{q-b^{m}}+\left( Nq-\overset{N}{%
\underset{m=1}{\sum }}\frac{q^{2}+1}{q-b^{m}}\right) a_{x}^{N-1}
\end{equation*}
or
\begin{equation*}
\overset{N}{\underset{m=1}{\sum }}\frac{b_{t}^{m}}{q-b^{m}}=a^{N-1}\overset{N%
}{\underset{m=1}{\sum }}b_{x}^{m}+a^{N-1}\overset{N}{\underset{m=1}{\sum }}%
\frac{b^{m}b_{x}^{m}}{q-b^{m}}-\left( \overset{N}{\underset{m=1}{\sum }}%
b^{m}+\overset{N}{\underset{m=1}{\sum }}\frac{(b^{m})^{2}+1}{q-b^{m}}\right)
a_{x}^{N-1};
\end{equation*}
which simplifies to
\begin{equation*}
\overset{N}{\underset{m=1}{\sum }}\frac{%
b_{t}^{m}+[(b^{m})^{2}+1]a_{x}^{N-1}-a^{N-1}b^{m}b_{x}^{m}}{q-b^{m}}=a^{N-1}%
\overset{N}{\underset{m=1}{\sum }}b_{x}^{m}-\overset{N}{\underset{m=1}{\sum }%
}b^{m}a_{x}^{N-1}.
\end{equation*}

3. Taking into account (\ref{suma}) we conclude that the r.h.s. of this
equation vanishes so splitting the partial fraction decomposition in the
l.h.s. of it w.r.t. $q$ we get precisely (\ref{eq:b_t}).


The variables $b^k$ are very convenient for explicit computation of
commuting flows and conservation laws for the system (\ref{eq:a_i}). Since
we will need in fact not only the conservation law densities (for example $p$
appearing in the left hand sides of equations like (\ref{eq:p_t})) but also
the fluxes --- expressions inside the $(\ldots )_x$ in the right hand sides
of such equations --- we will call \emph{conservation laws} the equalities (%
\ref{eq:p_t}) themselves.

Now we are ready to explain our method of deriving explicit formulas for
conservation laws of (\ref{eq:a_i}) using the Riemann surface associated to
this integrable hydrodynamic type system. One should note that $N$ component
semi-Hamiltonian hydrodynamic type system has infinitely many conservation
laws and commuting flows --- both families are parameterized by $N$
arbitrary functions of a single variable (see details in \cite{Tsar}). In
many interesting cases this functional dependence cannot be presented in
explicit form. Nevertheless $N$ infinite series of conservation laws and
commuting flows can be constructed, for instance, if the corresponding
equation of associated Riemann surface is known. As one can prove (cf. for
example \cite{MaksTsar}) for a vast class of semi-Hamiltonian systems such
series form the complete basis for the (infinite-dimensional) linear space
of all conservation laws (or commuting flows). Another techniques for
proving completeness can be found in \cite{classmech}.

\medskip

\noindent\textbf{$N$ infinite series of conservation laws for (\ref{eq:b_t})}
can be found in three steps:

\begin{enumerate}
\item Let us expand\footnote{%
This expansion is given by the so called Lagrange--B\"{u}rmann inversion
(see \cite{classmech} for an example of its use).} $q$ with respect to the
local parameter $\lambda$ at the vicinity of each root $b^{k}$:
\begin{equation}
q^{(k)}(\lambda )=b^{k}+\lambda q_{1}^{(k)}+\lambda ^{2}q_{2}^{(k)}+\lambda
^{3}q_{3}^{(k)}+ \ldots, \qquad \lambda \rightarrow 0.  \label{quk}
\end{equation}%
All coefficients $q_{m}^{(k)}$ can be found recursively, for instance%
\begin{equation*}
q_{1}^{(k)}=\frac{(1+(b^{k})^{2})^{N/2}}{\underset{m\neq k}{\prod }%
(b^{k}-b^{m})}.
\end{equation*}

\item Substitute the series $q^{(k)}(\lambda )$ into (\ref{eq:q2p}),
obtaining expansion
\begin{equation}
p^{(k)}(\lambda )=-\frac{q^{(k)}(\lambda )}{\sqrt{1+(q^{(k)}(\lambda ))^{2}}}%
=h^{k}+\lambda p_{1}^{(k)}+\lambda ^{2}p_{2}^{(k)}+\lambda
^{3}p_{3}^{(k)}+\ldots ,  \label{pek}
\end{equation}%
where 
\begin{equation}
h^{k}=-\frac{b^{k}}{\sqrt{1+(b^{k})^{2}}}.  \label{eq:h_k}
\end{equation}%
Then again all conservation law densities $p_{m}^{(k)}$ can be found
recursively, for instance%
\begin{equation}
p_{1}^{(k)}=-\frac{(1+(b^{k})^{2})^{(N-3)/2}}{\underset{m\neq k}{\prod }%
(b^{k}-b^{m})}.  \label{pik}
\end{equation}

\item Substitute $p^{(k)}(\lambda )$ into the generating function $p$ of
conservation laws (\ref{eq:p_t})
\begin{equation*}
(p^{(k)}(\lambda ))_{t}+\left( \sqrt{1-(p^{(k)}(\lambda ))^{2}}\overset{N}{%
\underset{n=1}{\sum }}b^{n}\right) _{x}=0  
\end{equation*}%
and expand both sides w.r.t. the parameter $\lambda \rightarrow 0$. Matching
the coefficients of the same powers of $\lambda$ one obtains $N$ infinite
series of conservation laws, namely
\begin{equation}  \label{eq:clN}
(h^{k})_{t}+\left( \sqrt{1-(h^{k})^{2}}\overset{N}{\underset{n=1}{\sum }}%
b^{n}\right) _{x}=0,\text{ \ }(p_{1}^{(k)})_{t}=\left( \frac{h^{k}p_{1}^{(k)}%
}{\sqrt{1-(h^{k})^{2}}}\overset{N}{\underset{n=1}{\sum }}b^{n}\right) _{x},
\ldots .
\end{equation}
\end{enumerate}

\textbf{Remark}: As a by-product we obtain the hydrodynamic type system (\ref%
{eq:a_i}), (\ref{eq:b_t}) in a symmetric conservative form
\begin{equation}
(h^{k})_{t}=\left( \sqrt{1-(h^{k})^{2}}\overset{N}{\underset{m=1}{\sum }}%
\frac{h^{m}}{\sqrt{1-(h^{m})^{2}}}\right) _{x},\text{ \ }k=1,\ldots ,N,
\label{conserv}
\end{equation}%
where we utilized the inverse point transformation (cf. (\ref{eq:q2p}), (\ref%
{eq:h_k}))
\begin{equation*}
b^{k}=-\frac{h^{k}}{\sqrt{1-(h^{k})^{2}}}.
\end{equation*}%
Its first $N$ conservation laws are (\ref{eq:clN}) expressed in variables $%
h^{k}$ using (\ref{eq:h_k}). 

Alongside with the constructed $N$ series of conservation laws one can
obtain another (incomplete) set of conservation laws usually called Kruskal
series. Its form is much simpler and is symmetric w.r.t. the variables $b^i$%
. We give Kruskal series below in Section~\ref{sec:chain}.

\section{Hydrodynamic Chains and Kruskal Series of Conservation Laws}

\label{sec:chain}

If we introduce new variables (called moments)
\begin{equation}
B^{k}=\frac{1}{k+1}\overset{N}{\underset{m=1}{\sum }}(b^{m})^{k+1},\text{ \ }%
k=0,1,\ldots ,  \label{moment}
\end{equation}%
then hydrodynamic type system (\ref{eq:b_t}) implies infinitely many
equations
\begin{equation}
B_{t}^{0}=(N+2B^{1})B_{x}^{0}-B^{0}B_{x}^{1},\text{ \ \ \ }%
B_{t}^{k}=(kB^{k-1}+(k+2)B^{k+1})B_{x}^{0}-B^{0}B_{x}^{k+1},\text{ \ }%
k=1,2,\ldots ;  \label{bichain}
\end{equation}%
just first $N$ of them are independent since due to (\ref{moment}) all
higher moments $B^{N},B^{N+1},$... are polynomial expressions of the first $N
$ moments so hydrodynamic type system (\ref{eq:b_t}) in moments $B^{k}$
takes the form
\begin{equation}
B_{t}^{0}=(N+2B^{1})B_{x}^{0}-B^{0}B_{x}^{1},\text{ \ \ }%
B_{t}^{k}=(kB^{k-1}+(k+2)B^{k+1})B_{x}^{0}-B^{0}B_{x}^{k+1},\text{ \ }%
k=1,2,...,N-2,  \label{bins}
\end{equation}%
\begin{equation*}
B_{t}^{N-1}=[(N-1)B^{N-2}+(N+1)B^{N}(\mathbf{B})]B_{x}^{0}-B^{0}(B^{N}(%
\mathbf{B}))_{x},
\end{equation*}%
where the dependence $B^{N}(\mathbf{B})=B^{N}(B^{0},\ldots ,B^{N-1})$ is be
found from (\ref{moment}) using the standard combinatorial results on
symmetric polynomials. For instance,

1. if $N=2$, then%
\begin{equation*}
B^{2}(\mathbf{B})=B^{0}B^{1}-\frac{1}{6}(B^{0})^{3};
\end{equation*}

2. if $N=3$, then%
\begin{equation*}
B^{3}(\mathbf{B})=B^{0}B^{2}+\frac{1}{2}(B^{1})^{2}-\frac{1}{2}%
(B^{0})^{2}B^{1}+\frac{1}{24}(B^{0})^{4};
\end{equation*}

3. if $N=4$, then%
\begin{equation*}
B^{4}(\mathbf{B})=B^{0}B^{3}+B^{1}B^{2}-\frac{1}{2}(B^{0})^{2}B^{2}-\frac{1}{%
2}B^{0}(B^{1})^{2}+\frac{1}{6}(B^{0})^{3}B^{1}-\frac{1}{120}(B^{0})^{5},...
\end{equation*}

Nevertheless, one can consider infinitely many equations (\ref{bichain})
without above restrictions. Similar infinite chains of quasilinear
first-order equations (called hydrodynamic chains) are very useful for
integration of various semi-Hamiltonian systems appearing in applications.
An overview of this approach can be found in \cite{MaxDorf, classmech}.

The associated Riemann surface (\ref{riman}) can be expanded at infinity as $%
q\rightarrow \infty $ and $\lambda \rightarrow 1$. For convenience we
replace below $\lambda $ by $\mu =-\ln \lambda $, so $\mu \rightarrow 0$ and
\begin{equation*}
\mu =\frac{N}{2}\ln (1+q^{2})-\overset{N}{\underset{k=1}{\sum }}\ln
(q-b^{k}).
\end{equation*}%
Then (see (\ref{moment})) asymptotically
\begin{equation}
\mu =\frac{N}{2}\ln \left( 1+\frac{1}{q^{2}}\right) +\overset{\infty }{%
\underset{k=0}{\sum }}\frac{B^{k}}{q^{k+1}}=\overset{\infty }{\underset{k=0}{%
\sum }}\frac{C^{k}}{q^{k+1}}, \qquad q\rightarrow \infty,  \label{ryad}
\end{equation}%
and introducing new field variables $C^k$ using their relation to $B^k$ in (%
\ref{ryad}) we obtain another remarkable hydrodynamic chain
\begin{equation}
C_{t}^{k}=(kC^{k-1}+(k+2)C^{k+1})C_{x}^{0}-C^{0}C_{x}^{k+1},\text{ \ }%
k=0,1,2, \ldots ,  \label{cichain}
\end{equation}%
where $C^{2k}=B^{2k}$ and $C^{2k-1}=B^{2k-1}-(-1)^{k}\frac{N}{2k}$.

Embedding of hydrodynamic type system (\ref{eq:b_t}) into the hydrodynamic
chain (\ref{cichain}) \emph{for arbitrary $N$} allows us to find Kruskal
conservation laws\footnote{%
We call this asymptotic expansion 
Kruskal, because M.~Kruskal was the first who introduced a similar
construction for the KdV equation.} in a compact form. First we (using the
Lagrange--B\"{u}rmann inversion) get from (\ref{ryad}) the asymptotic
decomposition of $q$ and $p$ as $\mu \rightarrow 0$:
\begin{equation}
q(\mu )=\frac{C^{0}}{\mu }+\frac{C^{1}}{C^{0}}+\mu \left( \frac{C^{2}}{%
(C^{0})^{2}}-\frac{(C^{1})^{2}}{(C^{0})^{3}}\right) +\mu ^{2}\left( \frac{%
C^{3}}{(C^{0})^{3}}-\frac{3C^{1}C^{2}}{(C^{0})^{4}}+\frac{2(C^{1})^{3}}{%
(C^{0})^{5}}\right) +\ldots ,  \label{qu_mu}
\end{equation}%
\begin{equation}
p(\mu )=-1+\frac{\mu ^{2}}{2(C^{0})^{2}}-\mu ^{3}\frac{C^{1}}{(C^{0})^{4}}%
+\mu ^{4}\left( -\frac{C^{2}}{(C^{0})^{5}}+\frac{5(C^{1})^{2}}{2(C^{0})^{6}}-%
\frac{3}{8(C^{0})^{4}}\right) +\ldots .  \label{pe_mu}
\end{equation}%
Substitute the last expansion into (\ref{eq:p_t}), note that $a^{N-1} = -
C^0 $ and equate the coefficients at equal powers of $\mu$ obtaining the
Kruskal series of conservation laws. The first few of them are:
\begin{equation}
((C^{0})^{-2})_{t}=[2C^{1}(C^{0})^{-2}]_{x},\text{ \ \ }%
[C^{1}(C^{0})^{-4}]_{t}=\left( \frac{2(C^{1})^{2}}{(C^{0})^{4}}-\frac{C^{2}}{%
(C^{0})^{3}}-\frac{1}{2(C^{0})^{2}}\right) _{x},  \label{ci}
\end{equation}%
\begin{equation*}
\left( \frac{C^{2}}{(C^{0})^{5}}-\frac{5(C^{1})^{2}}{2(C^{0})^{6}}+\frac{3}{%
8(C^{0})^{4}}\right) _{t}= \left( \frac{3C^{1}}{2(C^{0})^{4}} +\frac{%
5C^{2}C^{1}}{(C^{0})^{5}} -\frac{5(C^{1})^{3}}{(C^{0})^{6}} -\frac{C^{3}}{%
(C^{0})^{4}} \right) _{x}.
\end{equation*}

\section{Commuting Flows}

\label{sec:comm_flows}


The most important part of the integration procedure for a semi-Hamiltonian
system (as described briefly in Section~\ref{sec:integrability}) is
construction (preferably in an explicit form) of sufficiently many commuting
flows, either in a diagonal form (\ref{eq:comm_w}) or in non-diagonal
representation in terms of the variables $a^k$ or $b^k$. We choose the
latter possibility in this Section; the next Section is devoted to a
suitable modification of the Generalized Hodograph formula (\ref{eq:GHM}) to
the non-diagonal form of the commuting flows.

In the case considered in this paper the conservation laws for the
generating function have the form (see \cite{algebra})
\begin{equation*}
p_{t}=(S(p,b^{1},b^{2},\ldots ,b^{N}))_{x}
\end{equation*}%
where the flux $S(p,b^{1},b^{2},\ldots ,b^{N})$ can be represented in a much
simpler form $S(p,B^{0})$ (cf. \cite{maksgen}), where $B^{0}(\mathbf{b})$ is
the \textquotedblleft zeroth\textquotedblright\ moment (of the corresponding
hydrodynamic chain (\ref{bichain})). In this paper we consider the case $%
S=S_{1}(p,B^{0})=-B^{0}\sqrt{1-p^{2}}$. The situation with higher commuting
flows for the chain (\ref{bichain}) is as follows: the first commuting flow
has the flux (corresponding to the same generating function $p(\mathbf{b}%
,\lambda)$) of the form $S_{2}(p,B^{0},B^{1})$, the second commuting flow
has the flux $S_{3}(p,B^{0},B^{1},B^{2})$ etc. Assume that all these
functions $S_{1}(p,B^{0})$, $S_{2}(p,B^{0},B^{1})$, $%
S_{3}(p,B^{0},B^{1},B^{2})$, \ldots are the coefficients of an expansion of
some more complicated function w.r.t. an extra parameter $\zeta $, we come
to the ansatz that such a function should be written also in a compact form $%
G(p(\lambda ),p(\zeta ))$ (see a more general exposition in~\cite{MaxDorf}).

The conservation laws for the generating function of all higher commuting
flows were found in \cite{MaxDorf}. For this particular case they can be
represented formally as
\begin{equation}
\partial _{\tau (\eta )}p(\mu )=\partial _{x}G(p(\mu ),p(\eta )),
\label{gengen}
\end{equation}%
where $\mu =-\ln \lambda $, $p(\eta )$ is obtained by formally replacing the
parameter $\mu $ by the parameter $\eta$ and
\begin{equation}
G(p(\mu ),p(\eta ))=\frac{\sqrt{1-p^{2}(\mu )}}{\sqrt{1-p^{2}(\eta )}}+\frac{%
1}{2}p(\mu )\ln \frac{p(\eta )+1}{p(\eta )-1}+\ln \frac{p(\mu )-p(\eta )}{%
\sqrt{1-p^{2}(\mu )}+\sqrt{1-p^{2}(\eta )}}.  \label{gngn}
\end{equation}%
The so called \textquotedblleft vertex\textquotedblright\ operator $\partial
_{\tau (\eta )}$ is not yet determined and should be specified separately
for different cases below.

\subsection{The Kruskal Series of Commuting Flows}

For this series infinitely many respective fluxes for the generating
function of conservation laws of the higher commuting flows can be found
substituting the following asymptotic expansions for $\eta \rightarrow 0$
(cf. (\ref{qu_mu}) and (\ref{pe_mu}))
\begin{equation}
q(\eta )=\frac{C^{0}}{\eta }+\frac{C^{1}}{C^{0}}+\eta \left( \frac{C^{2}}{%
(C^{0})^{2}}-\frac{(C^{1})^{2}}{(C^{0})^{3}}\right) +\eta ^{2}\left( \frac{%
C^{3}}{(C^{0})^{3}}-\frac{3C^{1}C^{2}}{(C^{0})^{4}}+\frac{2(C^{1})^{3}}{%
(C^{0})^{5}}\right) +\ldots ,  \label{ququ}
\end{equation}%
\begin{equation}
p(\eta )=-1+\frac{\eta ^{2}}{2(C^{0})^{2}}-\eta ^{3}\frac{C^{1}}{(C^{0})^{4}}%
+\eta ^{4}\left( -\frac{C^{2}}{(C^{0})^{5}}+\frac{5(C^{1})^{2}}{5(C^{0})^{6}}%
-\frac{3}{8(C^{0})^{4}}\right) +\ldots  \label{pemu}
\end{equation}%
into (\ref{gngn}) and (\ref{gengen}) and specifying the expansion of the
vertex operator
\begin{equation*}
\partial _{\tau (\eta )}=\ln \eta \partial _{t^{0}}+\frac{1}{\eta }\partial
_{t^{1}}+\partial _{t^{2}}+\eta \partial _{t^{3}}+\eta ^{2}\partial
_{t^{4}}+\ldots 
\end{equation*}%
to match with the expansion
\begin{equation*}
G(p(\mu ),p(\eta ))=p(\mu )\ln \eta + \frac{C^{0}\sqrt{1-p^{2}(\mu )}}{\eta }
\end{equation*}%
\begin{equation*}
{}+\left[\frac{C^{1}}{C^{0}}\sqrt{1-p^{2}(\mu )}-p(\mu )\ln C^{0}+\frac{1}{2}%
\ln \frac{p(\mu )+1}{p(\mu )-1}-p(\mu )\ln 2\right]
\end{equation*}%
\begin{equation*}
+\eta \left[ \left( \frac{C^{2}}{(C^{0})^{2}}-\frac{(C^{1})^{2}}{(C^{0})^{3}}%
+\frac{1}{2C^{0}}\right) \sqrt{1-p^{2}(\mu )}-\frac{C^{1}}{(C^{0})^{2}}p(\mu
)-\frac{1}{C^{0}\sqrt{1-p^{2}(\mu )}}\right] +\ldots .
\end{equation*}%
So now we can identify $t^{0}\equiv x$ and $t^{1}\equiv t$, so
\begin{equation*}
(p(\mu ))_{t^{0}}=\left( p(\mu )\right) _{x}, \quad (p(\mu
))_{t^{1}}=-\left( C^{0}\sqrt{1-p^{2}(\mu )}\right) _{x},
\end{equation*}%
while higher conservation laws are (cf. (\ref{eq:p_t}) and (\ref{suma}))
\begin{equation*}
(p(\mu ))_{t^{2}}=\left( \frac{C^{1}}{C^{0}}\sqrt{1-p^{2}(\mu )}-p(\mu )\ln
C^{0}+\frac{1}{2}\ln \frac{p(\mu )+1}{p(\mu )-1}-p(\mu )\ln 2\right) _{x},
\end{equation*}%
\begin{equation*}
(p(\mu ))_{t^{3}}=\left[ \left( \frac{C^{2}}{(C^{0})^{2}}-\frac{(C^{1})^{2}}{%
(C^{0})^{3}}+\frac{1}{2C^{0}}\right) \sqrt{1-p^{2}(\mu )}-\frac{C^{1}}{%
(C^{0})^{2}}p(\mu )-\frac{1}{C^{0}\sqrt{1-p^{2}(\mu )}}\right] _{x},\ldots
\end{equation*}
Corresponding higher Liouville equations are associated with higher
commuting hydrodynamic chains. For instance, the first higher Liouville
equation%
\begin{equation*}
f_{t^{2}}=\left( q^{2}+\frac{C^{1}}{C^{0}}q-\ln C^{0}+1-\ln 2\right)
f_{x}+(q^{2}+1)f_{q}\left( q\ln C^{0}+\frac{C^{1}}{C^{0}}\right) _{x}
\end{equation*}%
is associated with the first higher hydrodynamic chain commuting with (\ref%
{cichain}):
\begin{equation*}
C_{t^{2}}^{k}=C_{x}^{k+2}+\frac{C^{1}}{C^{0}}C_{x}^{k+1}-\left( \ln
C^{0}-1+\ln 2\right) C_{x}^{k}-[(k+2)C^{k+1}+kC^{k-1}]\frac{C_{x}^{1}}{C^{0}}
\end{equation*}%
\begin{equation*}
+[C^{1}(kC^{k-1}+(k+2)C^{k+1})-C^{0}((k+1)C^{k}+(k+3)C^{k+2})]\frac{C_{x}^{0}%
}{(C^{0})^{2}}.
\end{equation*}

\subsection{$N$ Principal Series of Commuting Flows}

As we remarked above any $N$ component semi-Hamiltonian hydrodynamic type
system has infinitely many conservation laws and commuting flows
parameterized by $N$ arbitrary functions of a single variable (see detail in
\cite{Tsar}), but in many interesting cases only $N$ infinite series of
conservation laws and commuting flows can be constructed (cf. \cite%
{MaksTsar, classmech, Tsar}); they usually form a complete basis in the
respective linear spaces. Completeness of $N$ series constructed here is
discussed in Conclusion.

In this case we again (as in the previous sub-Section) start from (\ref%
{gengen}) and (\ref{gngn}).

The fluxes for the generating function of conservation laws of corresponding
higher commuting flows can be found from the equations written in a
conservative form (cf. (\ref{conserv}))%
\begin{equation}
(h_{i})_{\tau (\zeta )}=\partial _{x}G(h_{i},p(\zeta )).  \label{flows}
\end{equation}%
Substitution (see (\ref{pek})) of the expansion of $p(\zeta )$ as $\zeta
\rightarrow 0$ leads to construction of higher commuting flows. However,
this derivation is not so trivial. Indeed, corresponding expansion of the
\textquotedblleft vertex\textquotedblright\ operator $\partial _{\tau (\zeta
)}$ matching the obtained below expansion of the r.h.s. of (\ref{flows}) is
simple: 
\begin{equation}
\partial _{\tau ^{(k)}(\zeta )}=\partial _{t^{0,k}}+\zeta \partial
_{t^{1,k}}+\zeta ^{2}\partial _{t^{2,k}}+\zeta ^{3}\partial
_{t^{3,k}}+\ldots ,  \label{vertex}
\end{equation}%
where the index $k = 1, \ldots , N$ in $\tau ^{(k)}(\zeta )$ means $k$-th
branch of Riemann surface (\ref{riman}). However, direct substitution of (%
\ref{pek}) and (\ref{vertex}) yields the desirable infinite set of equations%
\begin{equation*}
(h^{i})_{t^{0,k}}=\partial _{x}G(h^{i},h^{k}),\text{ \ \ }%
(h^{i})_{t^{1,k}}=\partial _{x}G_{1}(h^{i},h^{k},p_{1}^{(k)}),\text{ \ }%
(h^{i})_{t^{2,k}}=\partial
_{x}G_{2}(h^{i},h^{k},p_{1}^{(k)},p_{2}^{(k)}),\ldots
\end{equation*}%
only for \textit{distinct} indices $i$ and $k$, because the function $G(x,y)$
has a singularity (see (\ref{gngn})) for $x=y$:
\begin{equation}
G(x,y)=Q(x,y)+\ln (x-y),  \label{expan}
\end{equation}%
with the non-singular part
\begin{equation*}
Q(x,y)=\frac{\sqrt{1-x^{2}}}{\sqrt{1-y^{2}}}+\frac{1}{2}x\ln \frac{y+1}{y-1}%
-\ln \left( \sqrt{1-x^{2}}+\sqrt{1-y^{2}}\right) .
\end{equation*}%
To complete the construction for $i = k$, one should observe that
substitution of (\ref{pek}) into (\ref{expan}) leads to the asymptotic
expansion ($\zeta \rightarrow 0$)
\begin{equation*}
G(h^{k},p^{(k)}(\zeta ))=Q(h^{k},h^{k}+\zeta p_{1}^{(k)}+\zeta
^{2}p_{2}^{(k)}+\zeta ^{3}p_{3}^{(k)}+\ldots )+\ln (p_{1}^{(k)}+\zeta
p_{2}^{(k)}+\zeta ^{2}p_{3}^{(k)}+\ldots )+\ln \zeta .
\end{equation*}%
Since the leading term $\ln \zeta $ disappears in (\ref{flows}) after
differentiation, we obtain the necessary matching expansions for $G$ and $%
\partial _{\tau (\zeta )}$ for $i=k$. For instance,
\begin{equation*}
(h^{k})_{t^{0,k}}=[Q(h^{k},h^{k})+\ln p_{1}^{(k)}]_{x},\text{ \ }%
(h^{k})_{t^{1,k}}=\left( p_{1}^{(k)}\left.\frac{\partial Q(h^{k},p)}{%
\partial p}\right|_{p=h^{k}}+\frac{p_{2}^{(k)}}{p_{1}^{(k)}}\right)
_{x},\ldots
\end{equation*}%
Conservation law densities $p_{m}^{(k)}$ were described in 
Section~\ref{sec:canon}. Thus, $N$ first commuting flows are%
\begin{equation*}
(h^{i})_{t^{0,k}}=\left( \frac{\sqrt{1-(h^{i})^{2}}}{\sqrt{1-(h^{k})^{2}}}+%
\frac{1}{2}h^{i}\ln \frac{h^{k}+1}{h^{k}-1}+\ln \frac{h^{i}-h^{k}}{\sqrt{%
1-(h^{i})^{2}}+\sqrt{1-(h^{k})^{2}}}\right) _{x},\text{ \ }i\neq k;
\end{equation*}%
\begin{equation*}
(h^{k})_{t^{0,k}}=\left[ \frac{1}{2}h^{k}\ln \frac{h^{k}+1}{h^{k}-1}+\ln
\frac{\sqrt{1-(h^{k})^{2}}\underset{m\neq k}{\prod }\sqrt{1-(h^{m})^{2}}}{%
\underset{m\neq k}{\prod }(h^{k}\sqrt{1-(h^{m})^{2}}-h^{m}\sqrt{1-(h^{k})^{2}%
})}\right] _{x}.
\end{equation*}

\section{Generalized Hodograph Method}

\label{sec:GenHod}

In this Section we finish our procedure for integration of hydrodynamic type
system (\ref{eq:a_i}), also written in equivalent forms (\ref{velo}), (\ref%
{eq:b_t}), (\ref{conserv}), (\ref{bins}).

According to the Generalized Hodograph Method (see detail in \cite{Tsar}),
any 
generic solution $r^{i}(x,t)$ of a semi-Hamiltonian diagonal hydrodynamic
type system (\ref{eq:diag_v}) in a neighborhood of a generic point is given
in an implicit form by the algebraic system (\ref{eq:GHM}) for the unknowns $%
r^{i}(x,t)$, 
where $w^{i}(\mathbf{r})$ are the velocities of a generic commuting flow (%
\ref{eq:comm_w}). In arbitrary hydrodynamic variables $u^{i}(\mathbf{r})$
one can easily rewrite the algebraic system (\ref{eq:GHM}) (cf. \cite{Tsar})
as
\begin{equation}
x\delta _{k}^{i}-tv_{j}^{i}(\mathbf{u})=w_{j}^{i}(\mathbf{u}),  \label{matr}
\end{equation}%
where the hydrodynamic type system (\ref{eq:diag_v}) has the form%
\begin{equation*}
u_{t}^{i}=\sum_{j}v_{j}^{i}(\mathbf{u})u_{x}^{j},\text{ \ }i,j=1,\ldots ,N,
\end{equation*}%
while commuting hydrodynamic type systems (\ref{eq:comm_w}) have the form
\begin{equation*}
u_{\tau }^{i}=\sum_{j}w_{j}^{i}(\mathbf{u})u_{x}^{j},\text{ \ }i,j=1,\ldots
,N.
\end{equation*}%
In this Section we will modify (\ref{eq:GHM}), (\ref{matr}) further in order
to get the simplest form suitable for the case studied.

In order to construct solutions of (\ref{eq:b_t}) we first need to prove the
following result suitable for our particular case of system (\ref{eq:a_i})
(cf.~\cite{MaksTsar, classmech}):

\textbf{Lemma}: \textit{Hydrodynamic type system} (\ref{eq:b_t}) \textit{%
together with commuting flows} (\ref{flows}) \textit{has the common
conservation law}%
\begin{equation*}
dz=\frac{1}{(C^{0})^{2}}dx+\frac{2C^{1}}{(C^{0})^{2}}dt+\left( \frac{1}{%
2(C^{0})^{2}}\ln \frac{p(\eta )+1}{p(\eta )-1}+\frac{p(\eta )}{%
(C^{0})^{2}(1-p^{2}(\eta ))}\right) d\tau (\eta ).
\end{equation*}

\textbf{Proof}: If $\mu \rightarrow 0$,
\begin{equation*}
G(p(\mu ),p(\eta ))=\mu ^{2}\left( \frac{1}{4(C^{0})^{2}}\ln \frac{p(\eta )+1%
}{p(\eta )-1}+\frac{p(\eta )}{2(C^{0})^{2}(1-p^{2}(\eta ))}\right) +\mu
^{3}(...)+...
\end{equation*}%
Taking into account (\ref{gengen}) and (\ref{pemu}), one can obtain%
\begin{equation*}
\left( \frac{1}{(C^{0})^{2}}\right) _{\tau (\eta )}=\left( \frac{1}{%
2(C^{0})^{2}}\ln \frac{p(\eta )+1}{p(\eta )-1}+\frac{p(\eta )}{%
(C^{0})^{2}(1-p^{2}(\eta ))}\right) _{x}.
\end{equation*}%
This conservation law together with (\ref{ci}) can be written in the above
potential form, where $z$ is a potential function such that $%
z_{x}=(C^{0})^{-2}$. The Lemma is proved.

Next we modify the initial Generalized Hodograph formula (\ref{eq:GHM}). It
should be reduced (following the idea in \cite{classmech}) to a form
suitable for our non-diagonal systems (\ref{eq:GHM}) and (\ref{eq:b_t}).

Algebraic system (\ref{eq:GHM}), (or (\ref{matr}) in arbitrary variables)
can be written in the form (here $\partial _{i}=\partial /\partial r^{i}$)
\begin{equation}
x+t\frac{\partial _{i}G_{1}}{\partial _{i}H_{0}}=\frac{\partial
_{i}G_{\infty }}{\partial _{i}H_{0}},  \label{temp}
\end{equation}%
where we denoted $H_{0}=(C^{0})^{-2},G_{1}=2C^{1}(C^{0})^{-2}$ and%
\begin{equation*}
G_{\infty }=\frac{1}{2(C^{0})^{2}}\ln \frac{p(\lambda )+1}{p(\lambda )-1}+%
\frac{p(\lambda )}{(C^{0})^{2}(1-p^{2}(\lambda ))}.
\end{equation*}
Here we returned to the original parameter $\lambda$ to indicate that now we
are working not with any particular asymptotic expansion (cf. (\ref{pek}), (%
\ref{pemu})), but with the original algebraic surface as a whole. Indeed,
hydrodynamic type system (\ref{eq:diag_v}) has a conservation law $\partial
_{t}H_{0}=\partial _{x}G_{1}$, while the commuting hydrodynamic system has
the conservation law $\partial _{\tau }H_{0}=\partial _{x}G_{\infty }$. This
means that $\sum\partial _{i}H_{0}\cdot r_{t}^{i}=\sum\partial
_{i}G_{1}\cdot r_{x}^{i}$ and $\sum\partial _{i}H_{0}\cdot r_{\tau
}^{i}=\sum\partial _{i}G_{\infty }\cdot r_{x}^{i}$. Taking into account (\ref%
{eq:diag_v}), (\ref{eq:comm_w}), (\ref{eq:GHM}) and splitting w.r.t. $%
r_{x}^{i}$, one obtains (\ref{temp}). Multiplying (\ref{temp}) by $\partial
_{i}H_{0}\cdot dr^{i}$ and summing up, one arrives at%
\begin{equation*}
xdH_{0}(\mathbf{r})+tdG_{1}(\mathbf{r}) = dG_{\infty }(\mathbf{r}).
\end{equation*}
Now we rewrite this equation after the invertible point transformation $(%
\mathbf{r})\rightarrow (\mathbf{b})$ as $xdH_{0}(\mathbf{b})+tdG_{1}(\mathbf{%
b})=dG_{\infty }(\mathbf{b})$, so the algebraic system (\ref{eq:GHM}) becomes%
\begin{equation*}
x\frac{\partial H_{0}}{\partial b^{i}}+t\frac{\partial G_{1}}{\partial b^{i}}%
=\frac{\partial G_{\infty }}{\partial b^{i}}.  
\end{equation*}%
Taking into account $\partial _{i}H_{0}=-2(C^{0})^{-3},\partial
_{i}G_{1}=2(C^{0})^{-2}b^{i}-4C^{1}(C^{0})^{-3}$ (see (\ref{moment}), here $%
\partial _{i}=\partial /\partial b^{i}$), we obtain the algebraic system
\begin{equation}  \label{eq:GHD_new}
x+t(2C^{1}-C^{0}b^{i})=\ln \left( \sqrt{1+q^{2}}-q\right) -q\sqrt{1+q^{2}}%
+C^{0}\frac{\sqrt{1+q^{2}}}{q-b^{i}}\left( \overset{N}{\underset{m=1}{\sum }}%
\frac{1}{q-b^{m}}-\frac{Nq}{1+q^{2}}\right) ^{-1},
\end{equation}%
where $q(\mathbf{b},\lambda )$ is the inverse function to the function $%
\lambda (\mathbf{b},q)$ 
(\ref{riman}). These $N$ equations are nothing but the diagonal part of the
matrix algebraic system (\ref{matr}). All off-diagonal equations are
compatible with the diagonal part (\cite{Tsar}).

So we proved:

\begin{theorem}
\label{theor1} Hydrodynamic type system (\ref{eq:b_t}) has infinitely many
particular solutions $b^{i}(x,t)$ in the implicit form given by (\ref%
{eq:GHD_new}) with a free parameter $\lambda$.
\end{theorem}

The algebraic system (\ref{eq:GHD_new}) determines one parametric family of
solutions $b^{i}(x,t,\lambda )$ in implicit form and simultaneously $%
g(x,t,\lambda )=-C^{0}(\mathbf{b}(x,t,\lambda ))$. Thus we found one
parametric family of Hamilton's equations (\ref{hamilton}), which are
Liouville integrable.

In fact, using the Generalized Hodograph Method and the \emph{nonlinear
superposition principle} implied by this method (see below) we easily obtain
multiparametric families of integrable metrics. Namely expanding the
generating function $p(\mathbf{b},\lambda )$ at different points on the
Riemann surface $p=p(\mathbf{b},\lambda )$ with the parameters $(p,\lambda )$
(for example when $p\rightarrow -1$ or $p\rightarrow b^{i}$), one can
construct infinite multiparametric series of new solutions $g(\mathbf{b}%
(x,t))$. Now we demonstrate this idea. 
To avoid large repeated expressions, we introduce functions%
\begin{equation*}
W_{i}(\mathbf{b},\lambda )=\ln \left( \sqrt{1+q^{2}}-q\right) -q\sqrt{1+q^{2}%
}+C^{0}\frac{\sqrt{1+q^{2}}}{q-b^{i}}\left( \overset{N}{\underset{m=1}{\sum }%
}\frac{1}{q-b^{m}}-\frac{Nq}{1+q^{2}}\right) ^{-1}.
\end{equation*}

\textbf{1}. \textit{Kruskal series}. Substitution of asymptotic expansion (%
\ref{ququ}) into (\ref{eq:GHD_new}) leads to%
\begin{equation*}
W_{i}(\mathbf{b},\mu )=\ln \mu +\frac{1}{\mu }W_{i}^{(-1)}(\mathbf{b}%
)+W_{i}^{(0)}(\mathbf{b})+\mu W_{i}^{(1)}(\mathbf{b})+\mu ^{2}W_{i}^{(2)}(%
\mathbf{b})+...,
\end{equation*}%
where, for instance,%
\begin{equation*}
W_{i}^{(-1)}(\mathbf{b})=C^{0}b^{i}-2C^{1}.
\end{equation*}
\begin{equation*}
W_{i}^{(0)}(\mathbf{b})=
(b^{i})^2 - \frac{C^{1}}{C^{0}} b^{i} +\frac{2(C^1)^2 - 3C^0 C^2}{(C^{0})^2}
 - \log C^0  - \log 2 .
\end{equation*}


\textbf{2}. $N$ \textit{principal series}.
Substitution of asymptotic series (\ref{quk}) into (\ref{eq:GHD_new}) leads
to%
\begin{equation*}
W_{i}^{(k)}(\mathbf{b},\lambda )=W_{i0}^{(k)}(\mathbf{b})+\lambda
W_{i1}^{(k)}(\mathbf{b})+\lambda ^{2}W_{i2}^{(k)}(\mathbf{b})+...,
\end{equation*}%
where, for instance,%
\begin{equation*}
W_{i0}^{(k)}(\mathbf{b})=\ln \left( \sqrt{1+(b^{k})^{2}}-b^{k}\right)
+(C^{0}\delta _{i}^{k}-b^{k})\sqrt{1+(b^{k})^{2}}.
\end{equation*}

Once we found all these coefficients $W_{im}^{k}(\mathbf{b})$ and the
Kruskal series $W_{i}^{k}(\mathbf{b})$, we can construct infinitely many
particular solutions parameterized by arbitrary number of constants $\sigma
_{k}^{m}$: 
\begin{equation}
x+t(2C^{1}-C^{0}b^{i})=\underset{k=1}{\overset{N}{\sum }}\underset{m=0}{%
\overset{\infty }{\sum }}\sigma _{k}^{m}W_{im}^{k}(\mathbf{b}),
\label{summa}
\end{equation}%
or a functional parameter $\varphi (\lambda )$:
\begin{equation}
x+t(2C^{1}-C^{0}b^{i})=\oint \varphi (\lambda )W_{i}(\mathbf{b},\lambda
)d\lambda ,  \label{inta}
\end{equation}%
where $\varphi (\lambda )$ and the contour can be chosen in many special
forms. Formulae (\ref{summa}) and (\ref{inta}) present \emph{the nonlinear
superposition principle} implied by the Generalized Hodograph Method.

\section{Conclusion}

\label{sec:final}

In this paper we considered integrability of semi-Hamiltonian hydrodynamic
type system (\ref{eq:a_i}). We found and presented $N$ infinite (principal)
series of conservation laws and commuting flows. This means that one can
extract infinitely many corresponding solutions by the Generalized Hodograph
Method. Here we did not investigated a completeness of conservation law
densities (and correspondingly commuting flows). This problem should be
investigated elsewhere. Here we just mention that (see (\ref{pik})) for $%
N=2n+1$, $n>0$
\begin{equation*}
\overset{2n+1}{\underset{k=1}{\sum }}p_{1}^{(k)}=-\frac{(1+(b^{k})^{2})^{n-1}%
}{\underset{m\neq k}{\prod }(b^{k}-b^{m})}=0.
\end{equation*}%
This means that corresponding conservation law densities $p_{1}^{(k)}$ are
not linearly independent for odd $N$. Thus one probably should construct
additional conservation law densities 
to get the complete set.

Semi-Hamiltonian hydrodynamic type system (\ref{eq:a_i}) holds for the
coefficients $a^{k}(x,t)$ of a first (polynomial) integral (\ref%
{firstintegral}). In this paper we found more appropriate set of field
variables $b^{k}$. This natural choice of unknown functions follows also
from factorization of polynomial expression for the first integral with
respect to ratio of both momenta\footnote{%
we remind that $a^{N}=1$ and $a^{N-1}=g$.}:

\begin{equation*}
F=\overset{N}{\underset{k=0}{\sum }}\frac{(-1)^k a^{k}}{g^{N-k}}%
p_{1}^{N-k}p_{2}^{k}= \left( \frac{p_{1}}{g}\right) ^{N}\overset{N}{\underset%
{k=1}{\prod }}\left( q 
-b^{k}\right) .
\end{equation*}%
Also this first integral can be written in two other equivalent forms (see (%
\ref{moment}) and (\ref{ryad})):
\begin{equation*}
F=(p_{2})^{N}\exp \left( -\overset{\infty }{\underset{k=0}{\sum }}\frac{%
B^{k}}{q^{k+1}}\right) =\left[ (p_{2})^{2}+\left( \frac{p_{1}}{B^{0}}\right)
^{2}\right] ^{N/2}\lambda (q,\mathbf{B}),
\end{equation*}%
where%
\begin{equation*}
q=\frac{B^{0}p_{2}}{p_{1}}.
\end{equation*}
These two equivalent representations for the first integral are of big
interest: if one finds another finite-dimensional parameterization of the
moments $B^{k}(\mathbf{b})$ compatible with (\ref{bichain}), the first
integral associated with this solution will be no longer polynomial. This
more general problem should be considered in a separate publication.

\section*{Acknowledgements}

M.V. Pavlov's work was partially supported by the grant of Presidium of RAS
\textquotedblleft Fundamental Problems of Nonlinear
Dynamics\textquotedblright\ and by the RFBR grant 15-01-01671-a. S.P. Tsarev
acknowledges partial financial support from the grant from Russian Ministry
of Education and Science to Siberian Federal University (contract $N^{o}$
1.1462.2014/K).


\addcontentsline{toc}{section}{References}

\begin{thebibliography}{99}
\bibitem{Bialy} \emph{M. Bialy, A. Mironov}, \newblock Rich quasi-linear
system for integrable geodesic flows on 2-torus, Discrete and Continuous
Dynamical Systems - Series A. 2011. V. 29. N. 1. P. 81-90. \emph{M. Bialy,
A. Mironov}, \newblock Cubic and quartic integrals for geodesic flow on
2-torus via system of hydrodynamic type, Nonlinearity. 2011. V. 24. P.
3541-3554. \emph{M. Bialy, A. Mironov}, \newblock Integrable geodesic flows
on 2-torus: Formal solutions and variational principle, Journal of Geometry
and Physics, Vol. 87, No. 1 (2015), P. 39-47.

\bibitem{BolMatFom} \emph{Bolsinov, A. V., Matveev, V. S., Fomenko, A. T.}
Two-dimensional Riemannian metrics with integrable geodesic flows. Local and
global geometry. Sbornik: Mathematics, 189(10), 1441-1466, 1998.

\bibitem{BolFom} \emph{Bolsinov A. V., Fomenko A. T.} Integrable Hamiltonian
systems: geometry, topology, classification. -- CRC Press, 2004.

\bibitem{dn} \emph{B.A. Dubrovin, S.P. Novikov,} \newblock Hamiltonian
formalism of one-dimensional systems of hydrodynamic type and the
Bogolyubov-Whitham averaging method, Soviet Math. Dokl., \textbf{27} (1983)
665--669. \emph{B.A. Dubrovin, S.P. Novikov,} \newblock Hydrodynamics of
weakly deformed soliton lattices. Differential geometry and Hamiltonian
theory, Russian Math. Surveys, \textbf{44} No. 6 (1989) 35--124.

\bibitem{deryabin1997} \emph{M. V. Deryabin}, \newblock Polynomial integrals
of dynamical systems and the Lax reduction, Math. Notes, \textbf{61} No. 3
(1997), 363--365.

\bibitem{GT} \emph{J. Gibbons, S.P. Tsarev}, \newblock Reductions of the
Benney equations, Phys. Lett. A \textbf{211} No. 1 (1996) 19-24. \emph{J.
Gibbons, S.P. Tsarev}, \newblock Conformal maps and reductions of the Benney
equations, Phys. Lett. A \textbf{258} No. 4-6 (1999) 263-271.

\bibitem{kozlov} \emph{V.V. Kozlov}, \newblock Polynomial integrals of
dynamical systems with one-and-a-half degrees of freedom. (Russian) Mat.
Zametki \textbf{45} No. 4 (1989) 46--52; translation in Math. Notes \textbf{%
45} No. 4 (1989) 296--300.

\bibitem{algebra} \emph{M.V. Pavlov}, \newblock Algebro-geometric approach
in the theory of integrable hydrodynamic type systems, Comm. Math. Phys.
\textbf{272} No. 2 (2007) 469-505.

\bibitem{maksgen} \emph{M.V. Pavlov}, \newblock Classification of integrable
hydrodynamic chains and generating functions of conservation laws, J. Phys.
A: Math. Gen., \textbf{39} No. 34 (2006) 10803-10819.

\bibitem{MaxDorf} \emph{M.V. Pavlov}, \newblock Integrable hydrodynamic
chains associated with Dorfman Poisson brackets. arXiv:1008.4530.

\bibitem{MaksTsar} \emph{M.V. Pavlov, S.P. Tsarev}, \newblock %
Tri-Hamiltonian structures of Egorov systems of hydrodynamic type, Funct.
Anal. Appl., \textbf{37} No. 1 (2003) 32-45.

\bibitem{classmech} \emph{M.V. Pavlov, S.P. Tsarev}, \newblock Classical
Mechanical Systems with one-and-a-half Degrees of Freedom and Vlasov Kinetic
Equation, in: \emph{Topology, Geometry, Integrable Systems, and Mathematical
Physics}, AMS Translations: Series 2, vol. 234, 2014, pp. 337-371.

\bibitem{RozYa} \emph{Rozhdestvensky B. L., Yanenko N. N.} \newblock
Systems of quasilinear equations and their applications to gas dynamics.
Translations of Mathematical Monographs, \textbf{55}. AMS, Providence, RI,
1983.

\bibitem{Tsar} \emph{S.P. Tsarev}, \newblock On Poisson brackets and
one-dimensional Hamiltonian systems of hydrodynamic type, Soviet Math.
Dokl., \textbf{31} (1985) 488--491. \emph{S.P. Tsarev}, \newblock The
geometry of Hamiltonian systems of hydrodynamic type. The generalized
hodograph method, Math. USSR Izvestiya, \textbf{37} No. 2 (1991) 397--419.
\end{thebibliography}
\end{document}